\documentclass[twocolumn,showpacs,preprintnumbers,amsmath,eqsecnum, amssymb,prd,superscriptaddress]{revtex4-1}

\usepackage{color}

\begin{document}
\preprint{CALT-TH 2016-010, IPMU16-0057, RUP-16-12}

\title{Bulk Local States and Crosscaps in Holographic CFT}


\author{Yu Nakayama}

\affiliation{Department of Physics, Rikkyo University, Toshima, Tokyo 175-8501, Japan}

\affiliation{Kavli Institute for the Physics and Mathematics of the Universe (WPI),
 University of Tokyo, 
 Kashiwa, Chiba 277-8583, Japan}

\author{Hirosi Ooguri}

\affiliation{Kavli Institute for the Physics and Mathematics of the Universe (WPI), University of Tokyo, 
 Kashiwa, Chiba 277-8583, Japan}
\affiliation{Walter Burke Institute for Theoretical Physics, 
California Institute of Technology, Pasadena, CA 91125, USA }
\affiliation{Center for Mathematical Sciences and Applications and 
Center for the Fundamental Laws of Nature, 
Harvard University, Cambridge, MA 02138, USA}

\begin{abstract}
In a weakly coupled gravity theory in the anti-de Sitter space,
local states in the bulk are 
linear superpositions of Ishibashi states for a crosscap
in the dual conformal field theory. 
The superposition structure can be constrained 
either by the microscopic causality in the bulk gravity or the
bootstrap condition in the boundary conformal field theory. 
We show, contrary to some expectation, that these two conditions
are not compatible to each other in the weak gravity regime. 
We also present an evidence to show that bulk local states in three
dimensions are not organized by the Virasoro symmetry. 

\end{abstract}

\maketitle 

\section{Introduction}

In our previous paper \cite{Nakayama:2015mva}, following \cite{Verlinde:2015qfa, Miyaji:2015fia}, we pointed out
that a bulk local state in a gravity theory in the anti-de Sitter space  (AdS) is 
a linear superposition of Ishibashi states \cite{Ishibashi:1988kg} for a crosscap
 in the dual conformal field theory (CFT). In this paper, we will discuss how to take the linear
superposition. 

Ishibashi's original construction is for boundary states, but they can be turned into crosscap states
by applying the dilatation by the imaginary unit, corresponding to translation 
by one quarter of the period in the global Lorentzian time in AdS. 
For each primary state $|\phi \rangle$, one can define
an Ishibashi state  $|\phi \rangle\rangle$,
\begin{align}
 M_{ab} |\phi \rangle\rangle = 0, ~~
 (P_a + K_a) |\phi \rangle \rangle = 0,
 \label{IshibashiCondition}
 \end{align}
preserving one half of 
the $SO(2, d)$ global conformal symmetry of 
${\mathbb R} \times S^{d-1}$, generated by the Hamiltonian $H$ along 
${\mathbb R}$, the rotation $M_{ab}$ of $S^{d-1}$, translation $P_a$ and
special conformal transformation $K_a$ ($a=1, ..., d$).
The equations (\ref{IshibashiCondition}) were solved explicitly
in \cite{Nakayama:2015mva} as,
\begin{align}
|\phi \rangle \rangle = \Gamma\left(\Delta-\frac{d}{2}+1\right)\left(\frac{P}{2}\right)^{d/2-\Delta} J_{\Delta-d/2}(P)|\phi \rangle , \label{momentum}
\end{align}
where $J_{\Delta-d/2}(P)$ is the Bessel function of the first kind with $\Delta$ being the scaling dimension of $\phi$.
It was observed in \cite{Nakayama:2015mva} that the dependence on the momentum
$P$ is the same as that for the bulk-boundary smearing function in AdS,
where the bulk point is evaluated at the center of AdS$_{d+1}$
\cite{Banks:1998dd, Balasubramanian:1998sn, Bena:1999jv}.
 (For simplicity, we are discussing Ishibashi states for scalar primaries. 
 See \cite{Nakayama:2015mva}
 for conditions when primary states carry non-zero spins.)

The question we would like to address is how to take a linear superposition
of Ishibashi states 
$|\phi\rangle\rangle$ over primary states $|\phi\rangle$
to construct a local state in the bulk AdS. 
One may be tempted to speculate that a bulk local state also
has a special role to play in the dual CFT. A natural guess would then
be that it satisfies consistency conditions for a crosscap in CFT,
in particular a bootstrap condition for crossing symmetry, 
which are analogous to the 
Cardy conditions on boundary states. 

We will show, contrary to such an expectation, that
the bootstrap condition in CFT contradicts with 
the microscopic causality in AdS, which has been proposed as conditions
on bulk local states in  
\cite{Hamilton:2006az, Kabat:2011rz, Kabat:2013wga, Kabat:2015swa,Kabat:2016zzr}. 
Namely, crosscap states obeying the 
 bootstrap constraints generically do not correspond to local states in the bulk.  
 We will also discuss bulk interpretation of crosscap states, which satisfy
 the bootstrap condition, and compare it with bulk local states satisfying the
 microscopic causality in AdS. 
 
When $d=2$, the conformal symmetry is enhanced to the Virasoro symmetry. 
 We will argue that a crosscap state in CFT preserves one 
 half of the Virasoro 
symmetry,  generalizing (\ref{IshibashiCondition}) to,
\begin{align}
  (L_n - (-1)^n \bar{L}_{-n}) | \phi\rangle\rangle_{\rm Virasoro}=0,
  \label{VirasoroIshibashi}
 \end{align}
for the left and right Virasoro generators, $L_n, \bar{L}_n$ ($n\in \mathbb{Z}$). On the other hand, 
we will present an evidence to show that
 the microscopic causality in AdS cannot be satisfied
by a linear superposition of Ishibashi states of the Virasoro symmetry
obeying (\ref{VirasoroIshibashi}). 
This also highlights the difference between local states in the bulk
and crosscap states on the boundary.

It would be desirable to understand how to characterize bulk local states in 
the language of CFT.  Our result shows that 
the bootstrap condition does not give a proper characterization of such states
and that the Virasoro symmetry in two dimensions does not give a useful 
guiding principle to solve the microscopic causality in AdS. 

This paper is  organized as follows. In section II, we will review
relations between the bulk and boundary coordinates and discuss
causality and crossing symmetry in these coordinates. 
In section III, we discuss the microscopic causality conditions for
local states in AdS and study solutions to these conditions. 
In section IV, we discuss the bootstrap condition on crosscaps in 
CFT, and compare their solutions to those of the microscopic causality
 conditions. In section V, we discuss a bulk interpretation of crosscap states.
 This also highlights the difference between crosscap states in CFT
 and local states in AdS. In section VI, we discuss whether crosscap states and bulk local states can be organized usefully in the AdS$_3$/CFT$_2$ case
 by Ishibashi states of the Virasoro symmetry.
We find that the answer is {\it yes} for crosscaps but {\it no} for bulk local states. 

\section{Causality and Cross-Ratio}

A crosscap state can be used to compute correlation functions of CFT 
on the real projective plane,
which is usually considered in the Euclidean signature. On the other hand, 
the causality in AdS should be discussed in the Lorentzian signature. Thus, in order
to compare the bootstrap condition on the projective plane 
and the microscopic causality in AdS, it is useful to understand analytic 
continuation between coordinates. In this paper,  we will use the global 
coordinates $(t, \rho, \Omega)$ of AdS
with the metric, 
\begin{align}
  ds^2 = -  {\rm cosh}^2 \rho \ dt^2 + d\rho^2 + {\rm sinh}^2 \rho\ d\Omega^2, \label{GlobalCoordinates}
  \end{align}
where coordinates on $S^{d-1}$ are denoted by $\Omega$, which
is identified with a unit vector in ${\mathbb R}^d$. 
As a consequence of working in the global patch, 
the causal interpretation of the crosscap cross-ratio 
$\eta$, defined below, is slightly different from that discussed in 
\cite{Hamilton:2006az, Kabat:2011rz, Kabat:2013wga, Kabat:2015swa,Kabat:2016zzr} 
in the Poincar\'e patch. We work in  the global patch as we find it more convenient
to compare the microscopic causality and the condition for the  crosscap bootstrap.

In the Euclidean signature, 
the global coordinates with the metric,  
\begin{align}
ds^2 = \cosh^2\rho d\tau^2 + \sinh^2 \rho d\Omega^2 + d\rho^2,
\label{EuclideanGlobal}
\end{align}
and the Poincar\'e coordinates with the metric, 
\begin{align}
ds^2 = \frac{dz^2}{z^2} + \frac{dx^2}{z^2} ,
\label{EuclideanPoincare}
\end{align}
with $x \in {\mathbb R}^d$ 
cover the same $(d+1)$-dimensional hyperbolic space. 
In particular the center of Euclidean AdS $\tau = 0, \rho=0$ corresponds to $z=1, x=0$.

On the boundary, the two coordinates are related to each other by 
the standard formula for the radial quantization, $x = e^\tau \Omega$. 
Thus, the involution,
\begin{align}
 x \to  \frac{x}{x^2}, 
 \label{CrosscapInvolution}
 \end{align}
to define the real projective plane is, 
 \begin{align}
  (\tau, \Omega) \to (-\tau, -\Omega). 
  \label{GlobalInvolution}
 \end{align}
In the following, an important role is played by the crosscap cross-ratio $\eta$
of two points $x_1$ and $x_2$ on the plane defined by,
\begin{align}
\eta = \frac{(x_1 - x_2)^2}{(1+x_1^2)(1+x_2^2)} \ .
\end{align}
In the Euclidean signature, we always have $0 \le \eta \le 1$, 
and $\eta =1$ corresponds to the limit in which $x_1$ approaches 
the image of $x_2$, namely $x_1 \to -x_2/ x_2^2$.
To see this, we note that $\eta \leq 1$ is equivalent to,
\begin{align}
1+ 2x_1\cdot x_2 + x_1^2 x_2^2 
= x_2^2 \left( \frac{x_2}{x_2^2} + x_1\right)^2 \ge 0 .
\end{align}

In the global coordinates (\ref{EuclideanGlobal}), the cross-ratio 
is expressed as,
\begin{align}
\eta = \frac{\cosh(\tau_1-\tau_2) - \Omega_1 \cdot \Omega_2}{\cosh(\tau_1-\tau_2) + \cosh(\tau_1+\tau_2)} ,  \label{dist}
\end{align}
Its Lorentzian continuation, $\tau = i t$, gives, 
\begin{align}
\eta = \frac{\cos(t_1-t_2) - \Omega_1 \cdot \Omega_2}{\cos(t_1-t_2) 
+ \cos(t_1+t_2)} ,  \label{dist}
\end{align}
and takes values in $ -\infty \le \eta \le \infty$. 
In the Lorentzian case,  $\eta=1$
corresponds to the limit where $(t_1, \Omega_1)$
and $(-t_2, -\Omega_2)$ are {\it light-like separated}. 
 
Let us discuss a bulk interpretation of $\eta$ and relate it to the causality.   Since the future light-cone from the center $(t = 0, \rho = 0)$ of AdS
reaches the boundary at $t= \pi/2$,
a boundary point $(t, \Omega)$ is space-like separated from the center
 if and only if  $ |t| < \pi/2$.  Using this fact, we can show that,
 when $\eta >1$, 
  at least  one pair of the three points are space-like 
 separated, modulo the $2\pi$ period in $t$.

\section{Microscopic Causality in the Bulk}

A bulk local operator $\hat\psi$ is a function (more generally a section)
over the bulk AdS and  acts on the Hilbert space of the dual CFT.  
In \cite{Nakayama:2015mva}, we required the action of the bulk isometry
on $\hat \psi$ to be compatible with that of the corresponding conformal symmetry in 
the CFT, 
\begin{align}
[J, \hat{\psi}] = i {\cal L}_{\mathcal{J}}  \hat{\psi}, 
\label{compatibility}
\end{align}
where 
$\mathcal{J}$ is a Killing vector of the AdS 
corresponding to any one of the conformal generators, $H, M_{ab}, P_a, K_a$,
and ${\cal L}_{\mathcal{J}}$ is the Lie derivative on $\hat \psi$ with respect to
$\mathcal{J}$.
Since the isotropy subgroup $SO(1,d)$ at the origin $(t=0, \rho=0)$ 
is generated by $M_{ab}$ and $P_a + K_a$, the bulk local operator there
should commute with them as, 
\begin{align}
[M_{ab},   \hat{\psi}(0)] &= 0  ,   \cr
[P_a + K_a, \hat{\psi}(0) ] &=  0  .
\label{commute}
\end{align}
Correspondingly, the state $|\psi(0)\rangle = \hat{\psi}(0) |0\rangle$ satisfies 
the condition, 
\begin{align}
M_{ab}|\psi(0) \rangle & = 0 ,   \nonumber \\
(P_a + K_a)|\psi(0) \rangle & =  0 ,
\end{align}
which we identified in \cite{Nakayama:2015mva}
as a condition for crosscap states in CFT.
Since Ishibashi states span the space of solutions to these equations, 
each bulk local state $|\psi(0) \rangle$ should be their linear superposition as, 
\begin{align}
  |\psi(0) \rangle = \sum_\phi \psi_\phi |\phi \rangle\rangle . 
\label{superposition}
\end{align}
The crosscap Ishibashi states may be regarded as a time-evolution of the boundary Ishibashi states by quarter period of the global Lorentzian time in AdS. 

If $\hat\psi(t,\rho,  \Omega)$  represents a single particle 
excitation in the bulk, it should approach  a single trace
primary operator $\phi_0(t, \Omega)$ at the boundary.
Thus, $\psi_{\phi_0}=1$ in (\ref{superposition}) and all other $\phi$ in the sum
should have scaling dimensions larger than that of $\phi_0$. 
If only $|\phi_0\rangle\rangle$ is in the sum, $\hat{\psi}$ would satisfy 
a free field equation in the bulk, because the crosscap 
Ishibashi state is an eigenstate of the Casimir operator of the conformal symmetry,  which is equal the Laplacian in AdS when
acting on $\hat \psi$ by (\ref{compatibility}).

To go beyond the free field limit in the bulk, it was proposed 
in \cite{Hamilton:2006az, Kabat:2011rz, Kabat:2013wga, Kabat:2015swa,Kabat:2016zzr} to
impose the microscopic causality:
\begin{align}
[\hat{\psi}(X), \hat{\psi}(Y)] = 0
\end{align}
when the two points, $X$ and $Y$ are space-like separated. 
It turns out that $\psi_\phi$ in the expansion (\ref{superposition})
can be determined order by order in the
large $N$ expansion, as demonstrated to order $1/N^2$ in \cite{Kabat:2016zzr}. 

The first non-trivial constraint coming from the microscopic causality is of
three-point functions (two on the boundary and one in the bulk). 
In the large $N$ limit, a bulk local state is equal to a particular Ishibashi state,
and  the three-point function can be expressed as
a two-point functions evaluated on the Ishibashi state.
 To write down the three-point function, 
it is convenient to use the Poincar\'e coordinates $(z, x)$ with the metric, 
\begin{equation}
ds^2 = \frac{dz^2 + dx^2}{z^2}.
\end{equation}
As shown in the previous section, the center of AdS in the global coordinates
corresponds to $(z=1, x=0)$, which is where the bulk point is evaluated in
the smearing function \eqref{momentum}.

The two-point function for primary fields, $\phi_1$ and $\phi_2$, at the 
boundary points $x_1, x_2$ evaluated on the Ishibashi state 
$|\phi_3 \rangle\rangle$ is given by, 
\begin{align}
& \langle 0 | \phi_1(x_1) \phi_2(x_2) 
 | \phi_3 \rangle\rangle \cr & =\frac{(1+x^2_1)^{\frac{\Delta_{2}-\Delta_1} {2}}  (1+x^2_2)^{\frac{\Delta_{1}-\Delta_2}{2}}}{ 
(x_1 - x_2)^{\Delta_1+\Delta_2} }\ g_{123}(\eta), 
\end{align}
where $\Delta_{1}$ and $\Delta_2$  
are scaling dimensions of the primary fields
dual to $\hat\psi_{1}$ and $\hat \psi_2$,
and 
\begin{align}
\eta= \frac{(x_1 -x _2)^2}{(1 + x_1^2)(1+x_2^2)},
\end{align}
is a cross-ratio invariant under the $SO(1,d)$ preserved by the crosscap. 
The function $g_{123}(\eta)$ takes the form,
\begin{align}
&~~ g_{123}(\eta) \cr
&= C_{123}  \ \eta^{\Delta_3/2} \cr
&\times {}_2F_1 \left(\frac{\Delta_{1}-\Delta_2 + \Delta_3}{2},\frac{\Delta_{2}-\Delta_1
+\Delta_3}{2}; \Delta_3 + 1-\frac{d}{2};\eta\right) , \cr \label{PCW}
\end{align}
where $C_{123}$ is the OPE coefficient of $\phi_1$ and $\phi_2$  into
$\phi_3$.
  
One way to derive \eqref{PCW} is to use the explicit form of the scalar OPE with conformal descendants \cite{McAvity:1995zd},
\begin{align}
&\phi_1(x_1) \phi_2(x_2) \cr
&= \sum_i \frac{C_{12i}}{(x_1-x_2)^{\Delta_1+\Delta_2-\Delta_i}} C^{\Delta_i,\Delta_1 - \Delta_2} (x_1-x_2, \partial_{x_2}) \phi_i(x_2) \cr
&~~ + \text{higher spin tensors} ,
\end{align}
where
\begin{align}
C^{a,b}(x,\partial) =& \frac{1}{B(a_+,a_-)} \int_0^1d\alpha \alpha^{a_+-1}(1-\alpha)^{a_- -1} \cr
&  \times \sum_{m=0} 
\frac{ \left(-\frac{1}{4}x^2 \alpha(1-\alpha) \partial^2\right)^m }{m! (a+1-\frac{1}{2}d)_m}e^{\alpha x\cdot \partial} ,
\end{align}
with $a_\pm = a\pm b$,  and evaluate the one-point functions
with the Ishibashi state, $|\phi_i \rangle\rangle$.

Alternatively, one may act the conformal Casimir on the two-point function 
and solve the eigenvalue problem in 
the OPE limit as a boundary condition of the second order differential equation. The latter approach is more or less equivalent to solving the Klein-Gordon equation in the AdS space-time from the holographic perspective \cite{Kabat:2011rz,Kabat:2013wga}. 
The three-point function computed in this prescription is the Wightman function, 
so the non-zero commutator outside of the lightcone generates as a cut 
in $\eta$ when 
any two of the three-points are light-like separated.

The function $g_{123}(\eta)$ for generic values of $\Delta_{1,2,3}$
has a cut in $\eta > 1$. As we saw in the previous section, 
when $\eta > 1$, at least one pair of the three points
becomes space-like separated. Therefore, 
it was proposed in \cite{Kabat:2015swa} that the cut in $\eta >1$
should be cancelled by superposition of Ishibashi state
and that this procedure determines the superposition 
coefficient $\psi_\phi$ order by
order in the $1/N$ expansions.

Let us illustrate the microscopic causality in AdS, by the following two examples.
First we consider a  free scalar field $\hat{\psi}$ in AdS. The three-point function of
two $\hat \psi$'s on the boundary and one composite operator
 $\hat \psi \times \hat\psi$ 
at the center of AdS, $z=1, x=0$, can be computed 
using the bulk boundary propagator, 
\begin{align}
\langle \hat \psi(z=0, x) \hat 
\psi(z=1, x=0)  \rangle = \left(\frac{1}{1+x^2}\right)^{\Delta}, 
\end{align}
as,
\begin{align}
&\left\langle \hat \psi (z=0, x_1) \hat
\psi(z=0, x_2) \left[\hat \psi\times\hat \psi\right] (z=1, x=0) 
\right\rangle_{\mathrm{AdS}}\cr
  &= \frac{2}{(1+x_1^2)^\Delta (1+x_2^2)^\Delta} .
\end{align}
Setting the right-hand side to be equal to $G(\eta)/(x_1-x_2)^{2\Delta}$,
we find,
\begin{align}
G(\eta) = 2\eta^\Delta,
\end{align}
which does not have singularity or cut at $\eta=1$, as expected. 

As another example, consider a local CFT in the bulk AdS (for holographic interpretation
of such a model, see \cite{Aharony:2015zea}). 
The three-point function can be computed using the conformal mapping 
from the flat space to AdS as, 
\begin{align}
&\langle \psi_1 (z=0, x_1) \psi_2(z=0, x_2) \psi_3(z=1, x=0) \rangle_{\mathrm{AdS}}  \cr
&= \frac{(1+x^2_1)^{\frac{\Delta_{2}-\Delta_{1}}{2}}   (1+x^2_2)^{\frac{\Delta_{1}-\Delta_{2}}{2}}}{ 
(x_1 - x_2)^{\Delta_1+\Delta_2} }G(\eta)
\end{align}
with
\begin{align}
G(\eta) = C_{123}\eta^{\frac{\Delta_3}{2}} .
\end{align}
Again we find no singularity or cut at $\eta=1$. 

In this case, we can compute the coefficient $\psi_{\phi}$ of the Ishibashi state
expansion (\ref{superposition}) of $|\hat\psi \rangle$ decomposition.
Assuming $\Delta_1 = \Delta_2$ for simplicity, we can expand $G(\eta)$ as
\begin{align}
\eta^{\frac{\Delta_3}{2}} 
 = &\sum_{n=0}\ C_n  \eta^{\frac{\Delta_3}{2}+n} \\
& \times
 {}_2F_1\left(\frac{\Delta_3}{2} + n, \frac{\Delta_3}{2}+n; \Delta_3 + 2n + 1-\frac{d}{2};\eta
 \right)\nonumber \ .
\end{align}
with
\begin{align}
C_n = \prod_{k=1}^n \frac{(\Delta_3+2k-2)^2}{2 k ( d-2(\Delta_3+n + k-1))}
\end{align}
Therefore, we need to add the infinite tower of Ishibashi states with even 
integer spacing to reproduce this bulk operator $\psi$. 
Note that the coefficients are not $1/N$ suppressed 
because the bulk theory is strongly interacting. Note also that not all the CFTs have such a structure of the operator spectrum.

\section{Bootstrap Condition on Crosscaps}

In this section, we consider
CFT on a $d$-dimensional real projective plane $\mathbb{RP}_d$, defined 
by quotienting
the flat Euclidean space ${\mathbb R}^d$ 
by the involution,
\begin{align}
x \to - \frac{x} {x^2} , \label{involution}
\end{align}
which preserves the $SO(1,d)$ subgroup of the 
Euclidean conformal symmetry $SO(1,d+1)$. 
The fundamental domain may be taken as $x^2 \ge 1$.

Conformally mapping the Euclidean space to the cylinder ${\mathbb R} \times S^{d-1}$, 
the involution \eqref{involution} becomes
$(\tau, \Omega) \to (-\tau,-\Omega)$,
where $\tau$ is a coordinate on $\mathbb R$, and 
$\Omega$
 is a unit vector in ${\mathbb R}^d$ 
parametrizing $S^{d-1}$. 
Analytically continuing to the Lorentzian signature cylinder,
$t = -i \tau$, the involution becomes $(t, \Omega) \rightarrow (-t, -\Omega)$
and the fundamental domain may be taken as $t \ge 0$.
If there is an additional global symmetry in CFT, 
the involution can be combined with $\phi \rightarrow \epsilon \phi$,
where $\epsilon$ is taken as a ${\mathbb Z}_2$ element of the 
symmetry so that the action is compatible with the OPE. 

Correlation functions of CFT on the real projective plane can be computed
by using the crosscap state, which is a superposition of the
Ishibashi states (\ref{momentum}) as, 
\begin{align}
|C\rangle = \sum_\phi A_\phi |\phi \rangle \rangle \ . \label{cardy}
\end{align}
The coefficient $A_\phi$  is related to the one-point function of a 
 primary operator $\phi$ on the projective plane:
\begin{align}
\langle \phi(x) \rangle_{\mathbb{RP}_d} 
= \frac{A_{\phi}}{(1+x^{2})^{\Delta_\phi}} , \label{onep}
\end{align}
where $\Delta_\phi$ is the scaling dimension of $\phi$. 
As noted n \cite{Nakayama:2015mva},
the rotational invariance demands that only scalar operators have non-zero
one-point functions on $\mathbb{RP}_d$.  

The two-point function of two scalar primary operators, 
$\phi_1(x_1)$ and $\phi_2(x_2)$, can be expressed as, 
\begin{align}
& \langle \phi_{1}(x_1) \phi_{2}(x_2) \rangle \cr
=& \frac{(1+x^2_1)^{\frac{\Delta_{2}-\Delta_1}{2}}   
(1+x^2_2)^{\frac{\Delta_{1}-\Delta_2}{2}}}{ (x_1 - x_2)^{\Delta_1+\Delta_2}}
\ G(\eta) \ ,
\end{align}
and $G(\eta)$ has the conformal partial wave decomposition as,
\begin{align}
&G(\eta)  = \sum_\phi C_{12\phi} A_\phi \eta^{\frac{\Delta_\phi}{2}} \label{SmallEta} 
 \cr
&~~ \times
 {}_2F_1\left(\frac{\Delta_1-\Delta_2+\Delta_\phi}{2}, \frac{-\Delta_1+\Delta_2+\Delta_\phi}{2}; 
 \Delta_\phi + 1-\frac{d}{2};\eta\right).    \cr 
\end{align}

Consistency of CFT on $\mathbb{RP}_d$ requires the crossing symmetry
of two-point functions \cite{Fioravanti:1993hf, Nakayama:2016cim}.
It compares the expansion (\ref{SmallEta}) at $\eta=0$ to another expansion
at $\eta=1$,  where 
$x_1$ approaches the mirror image of $x_2$. 
Since the OPE is convergent and the two-point functions are analytic, 
we obtain the crossing equation or crosscap conformal bootstrap equation,
\cite{Nakayama:2016cim},
\begin{align}
G(\eta) 
= \epsilon \left(\frac{\eta}{1-\eta}\right)^{\frac{\Delta_1 + \Delta_2}{2}} G(1-\eta) ,
\label{cross}
\end{align}
where the possibility of non-trivial involution $\epsilon$ was first introduced in \cite{Pradisi:1995qy}.

Clearly, both examples we discussed at the end of the last section --
the free massless scalar field and the local CFT in AdS, 
where $G(\eta) =C_{123} \eta^{\frac{\Delta_3}{2}}$ -- do not satisfy
 (\ref{cross}). This already shows a tension between the microscopic causality
and the bootstrap condition. 

To see that the bootstrap equation (\ref{cross}) is incompatible with
the bulk locality in general, we can apply the conformal partial wave decomposition
to the right-hand side of (\ref{cross}) as, 
\begin{align}
G(\eta) 
&= 
\epsilon \left(\frac{\eta}{1-\eta}\right)^{\frac{\Delta_1+\Delta_2}{2}} G(1-\eta) \cr
&=  \epsilon \eta^{\frac{\Delta_1+\Delta_2}{2}} \sum_\phi C_{12\phi} 
A_\phi (1-\eta)^{\frac{\Delta_\phi-\Delta_1-\Delta_2}{2}} \cr 
&~~\times  {}_2F_1 \left(\frac{\Delta_{1}-\Delta_2 + \Delta_\phi}{2},
\frac{-\Delta_1 +\Delta_{2}+\Delta_\phi}{2};\right. \cr 
& ~~~~~~~~~~~~~~ ~\left. \Delta_\phi + 1-\frac{d}{2};1- \eta\right) \label{rewriting} . 
\end{align}
We see that
$G(\eta)$ contains a cut at $\eta>1$ because of the  factor 
of $(1-\eta)^{\frac{\Delta_\phi-\Delta_1-\Delta_2}{2}}$
if $\Delta_\phi-\Delta_1 - \Delta_2$
is not an even integer.
We conclude that 
a solution to the bootstrap equation (\ref{cross}) cannot 
satisfy the microscopic causality, unless $C_{12\phi} A_\phi=0$
for all $\phi$'s with
$\Delta_\phi \notin \Delta_1+\Delta_2 + 2{\mathbb Z}$.

It may also be instructive to examine a  simple solution to the 
crosscap bootstrap equation, given by a free scalar field $\phi(x)$ 
in $d$-dimension with $\Delta_\phi = \frac{d}{2}-1$, and see if they
satisfy the microscopic causality.
By using the method of image, the two-point function on $\mathbb{RP}_d$ can be computed as
\begin{align}
&\langle \phi(x_1) \phi(x_2) \rangle_{\mathbb{RP}_d}\cr
&= \frac{1}{(x_1 - x_2)^{d-2}}+ \frac {\epsilon}{(1+2x_1\cdot x_2 + 
x_1^2 x_2^2)^{\frac{d}{2}-1}} \ ,
\end{align}
where $\epsilon = \pm 1$ reflects the additional $\mathbb{Z}_2$ symmetry on the free scalar field $\phi \rightarrow \pm \phi$, which can be combined with
the involution.
The corresponding $G(\eta)$ is given by,
\begin{align}
G(\eta) = 1+ \epsilon \left(\frac{\eta}{1-\eta} \right)^{\frac{d}{2}-1}. \label{freecross}
\end{align}
This satisfies the conformal bootstrap equation, but the microscopic causality
is violated when $d$ is not even.

\section{Gravity Dual of Crosscap States}

We found that the microscopic causality for local states in AdS and
the bootstrap condition for crosscap states in CFT are generically not 
compatible to each other. 
Given this, one may ask if crosscap states have a different
geometric interpretation in AdS.
 In this section, we discuss a straightforward interpretation using an 
 involution on AdS and find that its properties are different from those expected
 for bulk local states.

On the fixed AdS background, the involution acts on the global coordinates 
\begin{align}
ds^2 = -\cosh \rho^2 dt^2 + d\rho^2 + \sinh \rho^2 d\Omega^2 ,
\end{align}
as,
\begin{align}
(t, \rho, {\Omega}) \to (-t, \rho, -{\Omega}) . \label{bulkinvolution}
\end{align}
The involution preserves the $SO(2,d-1)$ subgroup of the AdS isometry, and at the boundary it reduces to the field theory involution discussed in the previous section. The bulk fields are identified as 
\begin{align}
\hat\psi(t,\rho, {\Omega}) \to \epsilon \hat\psi(-t,\rho,-{\Omega}) \ ,
\label{BulkCrosscap}
\end{align}
where we are allowing a possibility of an additional ${\mathbb Z}_2$ action $\epsilon$
on 
$\hat\psi$. 

After the Euclidean continuation, both the global coordinates $(\tau, \rho, \Omega)$
 and the Poincar\'e coordinates $(z, x)$ cover the entire hyperbolic space. 
Therefore, the Euclidean continuation of the
involution (\ref{BulkCrosscap}) can be expressed in the Poincar\'e coordinates
as, 
\begin{align}
(z,x) \to \left(\frac{z}{z^2+x^2}, \frac{-x}{z^2+x^2}\right) .
\end{align}
The fundamental domain can be taken $z^2 + x^2 \ge 1$.

The crosscap state $|C\rangle$ defined in this way is a superposition
of the Ishibashi states,
\begin{equation}
 |C\rangle = \sum_\phi A_\phi |\phi \rangle \rangle.
 \end{equation}
The coefficients $A_\phi$'s are computable in the bulk as
 the one-point function of the bulk field $\hat\psi$ dual to $\phi$ on the boundary.   
The one-point function vanishes unless $\phi$ is scalar. 

If the bulk gravity theory is weakly  coupled, the bulk
field $\hat\psi$ corresponding to a single-trace scalar operator $\phi$ 
can be described approximately by the free theory, 
\begin{align}
 S = \int d^{d+1}x \sqrt{-g} \left( \partial_\mu \hat\psi \partial^\mu \hat\psi
  + m^2 \hat\psi^2 \right).
  \label{freeaction} 
\end{align}
In this case, the one-point function vanishes because of the
$\mathbb{Z}_2$ symmetry of the action under $\psi \to - \psi$.
We therefore predict that, for all single trace operators,
$A_{\phi} = 0$ in the weakly coupled gravity regime. 
Note that this argument does not apply to multi-trace operators
since composites of even number of $\hat\psi$'s are $\mathbb{Z}_2$ even. 

Continuing to work in the weakly coupled gravity limit, 
two-point function of single trace operators $\phi_1$ and $\phi_2$
can be computed using
the method of image as,  
\begin{align}
G(\eta) = \delta_{\phi_1,\phi_2} 
\left(1 + \epsilon \left(\frac{\eta}{1-\eta} \right)^{\Delta_{\phi_1}} \right) , \label{genefree}
\end{align}
with the choice of the involution $\epsilon=\pm 1$.
This reproduces the two-point function of the generalized free field theory  
on the real projective plane and satisfies the crosscap bootstrap equation \eqref{cross}.  The conformal partial 
wave decomposition of (\ref{genefree}) generates infinite towers of double-trace operators as, 
\begin{align} 
&G(\eta) = \sum_n C_n \eta^{\Delta_{\phi_1} + n}\ \\
& \times {}_2F_1 \left(\Delta_{\phi_1} + n, \Delta_{\phi_1} + n;2\Delta_{\phi_1} 
+ 2n + 1-\frac{d}{2}; \eta\right), \nonumber
\end{align}
with $C_0 = 1, C_1= \epsilon \Delta_1, C_2 = \epsilon \frac{(-2+d)\Delta_1(1+\Delta_1)}{2(-6+d-4\Delta_1)}$ and so on. Therefore
the bulk crosscap state $| C\rangle$ contains the corresponding
 infinite towers of  Ishibashi states for the double trace operators 
 of the form, $\hat\psi \Box^n \hat\psi$
(This infinite sum can be truncated in the free massless scalar case since  
$\Box \hat\psi = 0$). 
This can be repeated for two-point functions of multi-trace operators to show
that the crosscap state $|C\rangle$ contains an infinite tower of multi-trace
Ishibashi states as well.

We have found that contributions of Ishibashi states for single-trace operators
are suppressed in crosscap states in the weakly coupled gravity limit
due to the $\mathbb{Z}_2$ symmetry $\hat \psi \rightarrow - \hat\psi$ 
of the free scalar action (\ref{freeaction}).
In contrast, bulk local states are dominated by single-trace states
in the same limit.  This also highlights the difference between crosscap states
and bulk local states. 

Recently it was suggested in \cite{Maloney:2016gsg} that 
CFT on the two-dimensional projective plane may not have a smooth geometric dual. 
This may be related to the fact the bulk involution \eqref{bulkinvolution}
 has a fixed point at the origin of AdS and quotieting by it may generate
 an orbifold singularity in the bulk.

\section{Enhancement to the Virasoro Symmetry}

When $d=2$, the global conformal symmetry is enhanced to the Virasoro symmetry. 
We will argue that crosscap states preserve one half of the full 
Virasoro symmetry in this
case and that we can use Ishibashi states for the full Virasoro symmetry
rather than the global conformal symmetry to expand crosscap states. 
On the other hand, we will provide some evidence that bulk local states are 
not necessarily
 organized by the Virasoro symmetry. 

\subsection{Virasoro Enhancement at Crosscaps}

Before discussing crosscap states, it would instructive to review the case for
boundary states. Consider a $(t, \sigma)$ plane and place a boundary located at $t=0$
and extending in the $\sigma$-direction.
Because of the scale invariance, 
the left and right-moving components of the  
energy-momentum tensor match at 
the boundary up to a total derivative along the boundary as, 
\begin{align}
T_{\sigma t}(t=0, \sigma) = T(t=0, \sigma) - \bar{T}(t=0, \sigma) = \partial_\sigma j_{\sigma\sigma}(\sigma) \ .
\end{align}
In addition, if we require the local Weyl invariance on the boundary,
the total derivative term must vanish $\partial_\sigma j_{\sigma\sigma}=0$ and that the boundary
preserves half of the bulk Virasoro symmetry 
 \cite{McAvity:1993ue}.
 
There is one more possibility: 
if we only require the global conformal invariance at the 
boundary,  the condition becomes \cite{Nakayama:2012ed},
\begin{align}
j_{\sigma\sigma} (\sigma)  = \partial_\sigma \ell_\sigma (\sigma) \ .
\end{align}
If this is non-zero, the boundary preserves one half
of the global conformal symmetry but not of the full Virasoro symmetry.
The reason why such a possibility exists at all is because we can always put an additional $0+1$ dimensional conformal quantum mechanical system at the boundary, which does not necessarily have the Virasoro symmetry.

The situation is different for crosscap states, where we cannot introduce 
localized degrees of freedom.
In particular, if the involution we used to define a crosscap acts trivially on 
the energy-momentum tensor,  the global conformal invariance alone demands
that the crosscap condition takes the form, 
\begin{align}
T(t,\sigma) - \bar{T}(-t,\sigma + \pi) = 0 .
\end{align}
In this case, one half of the full Virasoro symmetry is automatically preserved. 

One consequence of this is that the bootstrap condition with the Virasoro symmetry
 is the same as the one with only the global conformal symmetry. Indeed, 
 the numerical analysis in \cite{Nakayama:2016cim} shows that, in simple models
  such as the $2d$ critical Ising model, the bootstrap condition for the crosscap is so 
  strong that the solution automatically respects the Virasoro symmetry.

There is one caveat: When 
 the energy-momentum tensor is a part of a larger chiral algebra such as 
 the W-symmetry, there is a possibility to introduce non-trivial action 
 on $T_{\mu\nu}$ under the involution. From the holographic viewpoint, 
 this can happen in higher spin theories.

Preserving one-half of the Virasoro symmetry imposes  strong
constraints on solutions to the bootstrap equation. 
In fact, the constraints can be too strong to have any solution at all.
For example, a heterotic CFT with different values of 
Virasoro central charges for its left and right-movers
do not admit an involution on the real projective plane.

\subsection{No Virasoro Enhancement for Bulk Local States}

Let us turn to the microscopic causality conditions. 
We will use the $2d$ critical Ising model as an example to
see if the conditions can be satisfied by a superposition 
of Ishibashi states for the full Virasoro symmetry. 

In the critical Ising model, the Virasoro OPE gives, 
\begin{align}
[\sigma] \times [\sigma] = [1] + [\epsilon],
\end{align}
and because of the Virasoro symmetry, one may construct the crosscap state from the Virasoro Ishibashi states as,
\begin{align}
|C \rangle  = |1\rangle\rangle_{\mathrm{Virasoro}} + \frac{\sqrt{2}-1}{2} | \epsilon \rangle\rangle_{\mathrm{Virasoro}} . 
\end{align}
Correspondingly, the two-point function  of the spin operator $\sigma$ on the real projective plane can be decomposed as, 
\begin{align}
G_{\sigma\sigma}(\eta) &= (1-\eta)^{3/8} {}_2F_1
\left(\frac{3}{4},\frac{1}{4};\frac{1}{2};\eta\right) \\
 &+\frac{\sqrt{2}-1}{2} \eta^{1/2} (1-\eta)^{3/8} {}_2F_1
 \left(\frac{3}{4},\frac{5}{4};\frac{3}{2};\eta\right) .\nonumber
\end{align}
This two-point function satisfies the bootstrap equation for the crossing symmetry,
$G(\eta) = [\eta/(1-\eta)]^{\Delta_\sigma}G(1-\eta)$. 

Let us turn our attention to the microscopic causality condition. 
The question is whether it is possible to take an appropriate superposition,
\begin{align}
G(\eta) &= (1-\eta)^{3/8} {}_2F_1\left(\frac{3}{4},\frac{1}{4};\frac{1}{2};\eta\right) \cr
 &+\psi_{\epsilon}\ \eta^{1/2} (1-\eta)^{3/8} {}_2F_1\left(
 \frac{3}{4},\frac{5}{4};\frac{3}{2};\eta\right) ,
\end{align}
to cancel the cut at $\eta>1$ by adjusting the parameter $\psi_\epsilon$. It turns out that it is not possible. 
Since both conformal blocks,
\begin{align}
& {}_2F_1 \left(\frac{3}{4},\frac{5}{4};\frac{3}{2};\eta\right) 
 = \sqrt{\frac{2}{\eta}} \cdot \frac{\sqrt{1-\sqrt{1-\eta}}}{\sqrt{1-\eta}}  \cr
& {}_2F_1\left(\frac{3}{4},\frac{1}{4};\frac{1}{2};\eta\right)  = 
\frac{1}{\sqrt{2}} \cdot \frac{\sqrt{1+\sqrt{1-\eta}}}{\sqrt{1-\eta}} \ .
\end{align}
have cut for $\eta >1$ in both their denominators and numerators,
 it is not possible to cancel them by adjusting the single
 parameter $\psi_\epsilon$. 
In this case,  we cannot  construct a solution to the microscopic causality by a superposition of Ishibashi states for the Virasoro symmetry.

Though we do not expect that the $2d$ Ising model has a weakly coupled 
gravity description, this 
illustrates the difficulty in cancelling cuts at $\eta > 1$ 
by a superposition of Virasoro conformal blocks for a crosscap.

\section*{Acknowledgments}

We thank Alex Maloney, Tadashi Takayanagi, and Herman Verlinde for discussions. 
Our research is supported in part by the World Premier 
International Research Center Initiative (WPI Initiative), MEXT, Japan.
The research of HO is also supported in part by
U.S.\ DOE grant DE-SC0011632,
by the Simons Investigator Award, by Caltech's Walter Burke Institute for Theoretical Physics
and Moore Center for Theoretical Cosmology and Physics,
 by JSPS Grant-in-Aid for Scientific Research C-26400240, and 
 by JSPS Grant-in-Aid for Scientific Research on Innovative Areas 15H05895. 
HO thank the hospitality of the Institute for Advanced Study as Director's Visiting Professor in the fall 2015, where this work was initiated, 
and of the Center for Mathematical Sciences and
Applications and the Center for the Fundamental Laws of Nature at Harvard
University as a visiting scholar in the spring 2016,
where this work was completed.

\end{document}